\documentclass[twocolumn,showpacs,preprintnumbers,amsmath,amssymb,nofootinbib]{revtex4-1}
\usepackage{graphicx}% Include figure files
\usepackage{dcolumn}% Align table columns on decimal point
\usepackage{bm}% bold math
\usepackage{amsfonts}
\usepackage{mathrsfs}
\usepackage{comment}

\def\bequ{\begin{equation}}
\def\eequ{\end{equation}}
\def\be{\begin{equation}}
\def\ee{\end{equation}}

\begin{document}

%\preprint{APS/123-QED}

\title{Charged scalar and Dirac perturbations on a global monopole Reissner-Nordstr\"om-de Sitter black hole: quasinormal modes and strong cosmic censorship}

\author{Peiyang Li}
\author{Mengjie Wang}
\email{Corresponding author: mjwang@hunnu.edu.cn.}
%\email{Corresponding author: mjwang@hunnu.edu.cn.}
%
%\author{Qiyuan Pan}
%
%\email{panqiyuan@hunnu.edu.cn}
%
\author{Jiliang Jing}
%
%\email{jljing@hunnu.edu.cn}
%
%
\affiliation{\vspace{2mm}{
Department of Physics, Key Laboratory of Low Dimensional Quantum Structures and Quantum Control of Ministry of Education, Hunan Research Center of the Basic Discipline for Quantum Effects and Quantum Technologies, and Institute of Interdisciplinary Studies, Hunan Normal University, Changsha, Hunan 410081, P.R. China\vspace{1mm}}}
\date{\today}

\begin{abstract}
%We study perturbations of charged scalar and Dirac fields around Reissner-Nordstr\"om-de Sitter black holes with a global monopole. To do so, we first derive the equations of motion for both fields on the aforementioned background and these equations, for massive scalar fields with mass $\mu^2=\tfrac{2}{3}\Lambda$ and for massless Dirac fields, are reformulated uniformly in the Teukolsky formalism. By considering the fact that the Teukolsky equations in asymptotically de Sitter spacetimes may be mapped into the Heun equation, we are able to solve quasinormal spectra of scalar and Dirac fields by utilizing the Heun function method, not only for photon sphere modes but also for de Sitter and near-extremal modes. We analyze the spectra for both fields and all three types of modes by varying various parameters and, in particular, the global monopole $\eta^2$. In the near extremal regime, we further explore the role played by the global monopole and figure out that the presence of a global monopole, on one hand, leaves the strong cosmic censorship conjecture unaffected for scalar perturbations; while on the other hand, enhance the violation of the strong cosmic censorship conjecture for Dirac perturbations. Moreover, we identify that the impact of the global monopole both on the spectra and on the implications of SCC is performed by changing the multipole number. 
We study perturbations of charged scalar and Dirac fields around Reissner-Nordstr\"om-de Sitter black holes with a global monopole. To this end, we first derive the equations of motion for both fields on the aforementioned background; these equations are then reformulated uniformly into the Teukolsky equation. Since the Teukolsky equation in asymptotically de Sitter spacetimes can be mapped into the Heun equation, we are able to solve quasinormal spectra by employing the Heun function method, not only for photon sphere modes but also for de Sitter and near-extremal modes. We analyze the spectra of all three types for both fields and, in particular, ascertain the effects of the global monopole. In the near-extremal regime, we find that the presence of a global monopole, on the one hand, leaves the strong cosmic censorship conjecture unaffected for scalar perturbations, while on the other hand, it enhance the violation of strong cosmic censorship for Dirac perturbations. Furthermore, we identify that the impact of the global monopole on both the spectra and the strong cosmic censorship is achieved by a shift in the modified multipole number. Our work demonstrates that the Heun function method is an efficient and robust approach for exploring the interaction between asymptotically de Sitter black holes and perturbing fields. 
\end{abstract}

\maketitle

%%%%%%%%%%%%%%%%%
\section{Introduction}
%%%%%%%%%%%%%%%%%%%%%%%%%%%%%%%%%%%%%%%%%%%%%%%%%%%%%%%%%%%%%%%%%%%%%%%%%%%%%%
Black holes (BHs) are one of the most simplest macroscopic objects in the Universe, as suggested by the well-known no hair conjecture~\cite{Misner:1973prb}. They provide a unique arena for testing general relativity and exploring the interplay between classical gravity and quantum theory~\cite{Yagi:2016jml,Berti:2015itd,Bambi:2015kza}. In the era of gravitational waves, (binary) BHs are one of the most relevant gravitational wave sources~\cite{LIGOScientific:2016aoc,LIGOScientific:2016sjg,LIGOScientific:2017bnn,LIGOScientific:2020iuh,Mandel:2018hfr,LIGOScientific:2023lpe}. BH quasinormal modes (QNMs), in particular, plays a vital role in determining the ringdown phase of gravitational wave signals~\cite{Berti:2025hly,Barack:2018yly,Berti:2018vdi}. These modes are characterized by a discrete set of complex frequencies which only depend on the conserved charges of BHs and the mode indices of perturbation fields, and thus encode detailed information of the underlying spacetime geometry, see for example~\cite{Chandrasekhar:1985kt,Kokkotas:1999bd,Nollert:1999ji,Berti:2009kk,Konoplya:2011qq} and reference therein. 

BH QNMs have been explored extensively in the literature for various asymptotic spacetimes, and the asymptotically de Sitter (dS) spacetimes have been singled out from others for the following two reasons. On one side, the structure of quasinormal spectra in asymptotically dS BHs are much richer than the counterparts in asymptotically flat and asymptotically anti-de Sitter BHs. To be specific, there are \textit{three} qualitatively distinct families of modes emerged in asymptotically dS BHs, $i.e.$ the photon sphere (PS) modes which are associated with the potential barrier close to the unstable circular null geodesics, the dS modes which are governed mainly by the cosmological horizon, and the near-extremal (NE) modes which become relevant when the Cauchy horizon approaches the event horizon~\cite{Cardoso:2008bp,Du:2004jt,Zimmerman:2015trm,Cardoso:2017soq}. On the other side, the relative damping rates of different types of QNMs control the late time decay of linear perturbations and are closely related to the fate of the strong cosmic censorship (SCC) conjecture~\cite{Penrose:1969pc,Poisson:1989zz,Cardoso:2017soq}. In asymptotically dS spacetimes, the exterior decay of perturbations can compete with the interior blueshift in such a way that the Cauchy horizon remains sufficiently regular to allow weak extensions, thus SCC may be violated~\cite{Cardoso:2017soq}. The investigation of SCC has since been generalized to various settings, and a wealth of related works have emerged; for an incomplete studies, see for example~\cite{Dias:2018ynt,Dias:2018ufh,Mo:2018nnu,Destounis:2018qnb,Hod:2018dpx,Ge:2018vjq,Zhang:2019nye,Luna:2019olw,Gan:2019jac,Casals:2020uxa,Miguel:2020uln,Konoplya:2022kld,Shao:2023qlt,Cao:2024kht,Davey:2024xvd,Lin:2024beb} and references therein.

In order to look for quasinormal spectra, one normally has to employ numerical methods\footnote{There are also a few analytically solvable model, see for example~\cite{Bolokhov:2025rng} and references therein.}. The most commonly used approaches include the shooting method~\cite{Chandrasekhar:1975zza}, the WKB method~\cite{Schutz:1985km}, the continued fraction method~\cite{Leaver:1985ax}, the Horowitz-Hubeny method~\cite{Horowitz:1999jd}, the pseudospectral method~\cite{Wang:2019qja,Jansen:2017oag}, the matrix method~\cite{Lin:2016sch} and, for a further discussion on numerical methods, see for example in~\cite{Konoplya:2011qq} and references therein. Recently, the Heun function method has been introduced to investigate QNMs of BHs~\cite{Fiziev:2005ki,Hatsuda:2020sbn,Noda:2022zgk,Chen:2024rov,Chen:2025sbz,Xia:2025hwt}. The PS quasinormal spectra, in particular, have been explored for asymptotically dS spacetimes via the Heun function method. Then one may wonder if the Heun function method is applicable to solve the spectra of dS and NE modes for asymptotically dS BHs. As the first motivation of this paper, we explicitly show, by taking Reissner-Nordstr\"om-de Sitter (RN-dS) BHs with a \textit{global monopole} as an example, that this is \textit{indeed} the case. 

Topological defects may arise naturally in many particle physics models through spontaneous symmetry breaking in the early universe~\cite{Kibble:1976sj}. The global monopole is one of such topological defects arising from the spontaneous breaking of global $O(3)$ symmetry down to $U(1)$ symmetry~\cite{Barriola:1989hx}. When a global monopole is present in a spacetime, it introduces a deficit solid angle and induces various physical consequences in BH physics~\cite{Zhang:2023wwk,Wang:2021uix,Wang:2021upj,Pan:2008xz,Chen:2005vq,Piedra:2019tpx,Zhang:2014xha}. As the second motivation of this paper, we would like to illustrate the impact of the global monopole on the spectra of three distinct sets of modes and on SCC. 

The structure of this paper is organized as follows. In Section~\ref{seceq} we introduce the geometry of RN-dS BHs with a global monopole, and present equations of motion for charged scalar and Dirac fields on the aforementioned background. In Section~\ref{secnumerics} we briefly describe the Huen function method and the corresponding numerical results are presented in Section~\ref{secresults}. Final remarks and conclusions are presented in the last section. %Throughout this paper, we work in the geometric units with $c = G = 1$.

%%%%%%%%%%%%%%%%%%%%%%%%%%%%%%%%%%%%%
\section{background geometry and field equations}
\label{seceq}
%%%%%%%%%%%%%%%%%%%%%%%%%%%%%%%%%%%%%
In this section, we first briefly review the background geometry of RN-dS BHs with a global monopole, and then present equations of motion for charged scalar and Dirac fields on the aforementioned backgrounds in the Teukolsky formalism. 
\subsection{The line element}
The line element of a global monopole RN-dS BH may be written as~\cite{Gao:2002hf}
\begin{equation}
ds^2=-\dfrac{\Delta_r}{r^2}dt^2+\dfrac{r^2}{\Delta_r}dr^2+\tilde{\eta}^2r^2\left(d\theta^2+\sin^2\theta d\varphi^2\right) \;,\label{metric}
\end{equation}
where the metric function is given by
\begin{align}
\Delta_{r} &= r^2 \left( 1 - \frac{\Lambda}{3}r^2 \right)- 2Mr + Q^2 \;,\label{metricfunc1}
\end{align}
and the electromagnetic potential of the RN-dS BH is
\begin{equation}
  A_{\alpha} \, \mathrm{d}x^{\alpha} = -\frac{Q}{r}\mathrm{d}t\;.\label{eq:potential}
\end{equation}
Note that here $\Lambda$ is the cosmological constant, $M$ and $Q$ are the mass and charge parameters of the BH, and the dimensionless parameter $\tilde{\eta}^2$ is defined as
\begin{equation}
\tilde{\eta}^2 \equiv 1 - \eta^2 \;,\label{monopole}
\end{equation}
where $\eta$ is the global monopole parameter, and the RN-dS spacetimes may be recovered when $\eta^2=0$. It is shown clearly in Eq.~\eqref{metric} that a solid deficit angle is introduced due to the presence of the global monopole and the solid angle of this spacetime is $4\pi\tilde{\eta}^2$.

For later numerical convenience, the metric function given in Eq.~\eqref{metricfunc1} may be rewritten as  
\begin{align}
\Delta_{r} = -\frac{\Lambda}{3}(r - r_{+}^{\prime})(r - r_{+})(r - r_{-})(r - r_{-}^{\prime}) \;,\label{metricfunc}
\end{align}
where $r_+^\prime$, $r_+$, $r_-$ are respectively the cosmological horizon, the event horizon, the Cauchy horizon, and $r_-^\prime$ is given by
\begin{equation}
r_{-}^\prime=-(r_+^\prime+r_++r_-)\;.\nonumber
\end{equation}

The maximal charge of the BH is denoted by $Q_{max}$, indicating an extremal BH with $r_-=r_+$.
%%%%%%
\subsection{Equations of motion in the Teukolsky formalism}
%%%%%%
For our interests in this paper, in this subsection, we derive equations of motion for charged scalar and Dirac fields on a RN-dS BH with a global monopole. 

A massive charged scalar field $\Phi$ obeys%~\cite{Faraoni:1998qx}
\begin{equation}
  \left[
    \frac{1}{\sqrt{-g}} \mathcal{D}_{\alpha} \!\left( \sqrt{-g} \, g^{\alpha\beta} \mathcal{D}_{\beta} \right) - \mu^2 \right] \Phi(t,r,\theta,\varphi) = 0\;,
  \label{eqScalar}
\end{equation}
where $\mathcal{D}_{\alpha}\equiv\partial_{\alpha} - i e A_{\alpha}$, $A_\alpha$ is the electromagnetic potential given in Eq.~\eqref{eq:potential} and $e$ denotes the field charge. By taking the following ansatz for scalar fields
\begin{equation}
\Phi=e^{-i\omega t}e^{im \varphi}R(r)S(\theta)\;,\label{ScalarDecom}
\end{equation}
and by substituting Eqs.~\eqref{metric},~\eqref{eq:potential} and~\eqref{ScalarDecom} into Eq.~\eqref{eqScalar}, one immediately obtains the radial part of equations of motion for scalar fields %satisfying
\begin{equation}
\left[\dfrac{d}{dr}\left(\Delta_r\dfrac{d}{dr}\right)+\dfrac{K_r^2}{\Delta_r}-\mu^2r^2-\dfrac{\ell(\ell+1)}{\tilde{\eta}^2}\right]R(r)=0\;,\label{RadialS}
\end{equation}
with
\begin{equation}
K_r\equiv\omega r^2-eQr\;,\label{ExpK}
\end{equation}
where $\omega$ and $m$ are the frequency and azimuthal number of scalar fields, $\ell$ is the multipole number and $\ell=0, 1, 2, \cdots$.  

A massless charged Dirac field $\Psi$ satisfies
\begin{equation}
\gamma^\alpha(\mathcal{D}_\alpha-\Gamma_\alpha)\Psi=0\,,\label{ChargedDiraceqU1}
\end{equation}
where, similar to the scalar case, $\mathcal{D}_\alpha\equiv \partial_\alpha-ieA_\alpha$, and 
\begin{equation}
\Gamma_\alpha=-\frac{1}{8}(\gamma^a\gamma^b-\gamma^b\gamma^a)\Sigma_{ab\alpha}\,.\label{spincon}
\end{equation}
Here the matrices $\gamma^\alpha$ are associated to the geometry given in Eq.~\eqref{metric} through $\gamma^\alpha\gamma^\beta+\gamma^\beta\gamma^\alpha=2g^{\alpha\beta}$ so that one may construct $\gamma^\alpha$ as %may be constructed %are defined as~\cite{Wang:2017fie}
\begin{align}
&\gamma^t=\dfrac{r}{\sqrt{\Delta_r}}\gamma^0\,,\;\;\;\;\;\;\;\gamma^r=\dfrac{\sqrt{\Delta_r}}{r}\gamma^3\,,\nonumber\\
&\gamma^\theta=\dfrac{1}{\tilde{\eta}r}\gamma^1\,,\;\;\;\;\;\;\;\;\;\;\;\gamma^\varphi=\dfrac{1}{\tilde{\eta}r\sin\theta}\gamma^2\,,\label{gammams}
\end{align}
and
\begin{equation}
\Sigma_{ab\alpha}=e_a^\beta(\partial_\alpha e_{b\beta}-\Gamma^\rho_{\beta\alpha}e_{b\rho})\,,\nonumber
\end{equation}
where $\gamma^a(a=0, 1, 2, 3)$ are the ordinary Dirac matrices defined in the Bjorken-Drell representation~\cite{Bjorken:1965sts}, and the tetrad $e_a^\alpha$ may constructed in terms of $\gamma^\alpha=e_a^\alpha\gamma^a$. 

By taking the following ansatz for Dirac fields
\begin{equation}
\Psi=\left(
\begin{matrix}
\psi\\
\psi
\end{matrix}
\right)\,,\;\;\;\psi=\frac{e^{-i\omega t}e^{im\varphi}}{(r^2\Delta_r\sin^2\theta)^{1/4}}
\left(
\begin{matrix}
R_1(r)S_1(\theta)\\
R_2(r)S_2(\theta)
\end{matrix}
\right)\,,\label{Chargedfielddecom}
\end{equation}
one may obtain the radial part of equations of motion for Dirac fields directly 
\begin{align}
&\sqrt{\Delta_r}\dfrac{d}{dr}\left(\sqrt{\Delta_r}\dfrac{d}{dr}\right)R_{1}(r)+H_1(r)R_{1}(r)=0\,,\label{ChargedDiracU2radialR1}\\
&\sqrt{\Delta_r}\dfrac{d}{dr}\left(\sqrt{\Delta_r}\dfrac{d}{dr}\right)R_{2}(r)+H_2(r)R_{2}(r)=0\,,\label{ChargedDiracU2radialR2}
\end{align}
with
\begin{align}
&H_1(r)=\dfrac{K_r^2+\tfrac{i}{2}K_r\Delta_r^\prime}{\Delta_r}-iK_r^\prime-\dfrac{(\ell+\frac{1}{2})^2}{\tilde{\eta}^2}\,,\nonumber\\
&H_2(r)=\dfrac{K_r^2-\tfrac{i}{2}K_r\Delta_r^\prime}{\Delta_r}+iK_r^\prime-\dfrac{(\ell+\frac{1}{2})^2}{\tilde{\eta}^2}\,,\nonumber
\end{align}
where $K_r$ is defined in Eq.~\eqref{ExpK}, $\prime$ denotes a derivative with respect to $r$, $\ell$ is the multipole number of Dirac fields with $\ell=1/2, 3/2, \cdots$~\cite{Dolan:2009kj}.

By taking $\mu^2=\tfrac{2}{3}\Lambda$ and by letting $R_1=R_{-1/2}$, $R_2=\sqrt{\Delta_r}R_{1/2}$, Eqs.~\eqref{RadialS},~\eqref{ChargedDiracU2radialR1} and~\eqref{ChargedDiracU2radialR2} may be unified into the following Teukolsky equation
%%%%%%%%
\begin{equation}
\begin{aligned}
\Bigg[&\, \Delta_r^{-s}\dfrac{d}{d r}\!\left( \Delta_r^{\,s+1}\frac{d}{d r} \right)+ \frac{K_r^{2}-isK_r\,\Delta'_{r}}{\Delta_r}+2isK_r^\prime\\\quad&-\dfrac{2\Lambda}{3}(s+1)(2s+1) r^{2}+s(2s+1) - \tilde{\lambda}\Bigg] R_s(r)=0\;,\label{TeukolskyEq}
\end{aligned}
\end{equation}
where we take $s=0$ ($s=\pm1/2$) for scalar (Dirac) fields, $K_r$ is given in Eq.~\eqref{ExpK}, $\prime$ denotes a derivative with respect to $r$, and 
\begin{equation}
\tilde{\lambda}=\dfrac{\lambda+s(2s-1)}{\tilde{\eta}^2}\;,\;\;\;\;\;\;\lambda=\ell(\ell+1)-s(s-1)\;.
\end{equation}

When the monopole parameter is absent, Eq.~\eqref{TeukolskyEq} gives the standard Teukolsky equation of conformally coupled charged scalar and massless charged Dirac fields on RN-dS BHs.  

%%%%%%%%%%%%%%%%%%%%%%%%%%%%%%%%%%%%%
\section{Numerical Methods}
\label{secnumerics}
%%%%%%%%%%%%%%%%%%%%%%%%%%%%%%%%%%%%%
In this section, we first briefly introduce the (general) Heun equation and then solve Eq.~\eqref{TeukolskyEq} via the Heun function.

The (general) Heun equation is a second order Fuchsian differential equation with four regular singular points located at $z = 0, 1, a$, $\infty$. It is a natural extension of the Riemann hypergeometric equation and may be written as~\cite{Ronveaux1995}
\begin{equation}
\frac{d^{2}y}{dz^{2}}+\!\left(\frac{\gamma}{z}+\frac{\delta}{z-1}+\frac{\epsilon}{z-a}\right)\!\frac{dy}{dz}+\frac{\alpha\beta\, z - q}{z(z-1)(z-a)}\,y=0\;,\label{eq:Heun}
\end{equation}
where $a\neq 0, 1$, the Fuchs relation is satisfied
\begin{equation}
\gamma+\delta+\epsilon=\alpha+\beta+1\;,\label{eq:Fuchs-relation}
\end{equation}
and the \emph{accessory} parameter $q$ is related to the separation constant $\tilde{\lambda}$ in our problem.

In order to employ the Heun function method, one shall transform Eq.~\eqref{TeukolskyEq} into Eq.~\eqref{eq:Heun}. To do so, we first make a M\"obius transformation
\begin{equation}
z \;=\;\dfrac{(r_{+}'-r_{-})(r-r_{+})}{(r_{+}'-r_{+})(r-r_{-})}\;,\label{eqmobius}
\end{equation}
which brings the five singular points $(r_{+}',\, r_{+},\, r_{-},\, r_{-}',\, \infty)$ appeared in Eq.~\eqref{TeukolskyEq} into $(1,\, 0,\, \infty,\, z_{r},\, z_{\infty})$, where
\begin{equation}
z_{r}=\frac{(r_{+}'-r_{-})(r_{-}'-r_{+})}{(r_{+}'-r_{+})(r_{-}'-r_{-})}\;,\;\;\;\;\;\;z_{\infty}=\frac{r_{+}'-r_{-}}{\,r_{+}'-r_{+}\,}\;.\label{eqzr}
\end{equation}
We then further take the following transformation for the radial function
\begin{equation}
R_s(r)=z^{B_{1}}\,(z-1)^{B_{2}}\,(z-z_{r})^{B_{3}}\,(z-z_{\infty})^{\,2s+1}\,y(z)\;,\label{eqfactor}
\end{equation}
with
\begin{equation}
B_1=B(r_+)\;,B_2=B(r_+^\prime)\;,B_3=B(r_-^\prime)\;,B_4=B(r_-)\;,\nonumber%\label{eqBvalues}
\end{equation}
and where
\begin{equation}
B(r) \;=\; \dfrac{iK_r^\prime}{\Delta_{r}}\;.\label{eqBdef}
\end{equation}

With Eqs.~\eqref{eqmobius}--\eqref{eqBdef} at hand, now Eq.~\eqref{TeukolskyEq} turns into the (general) Heun equation given in Eq.~\eqref{eq:Heun} and where
\begin{align}
&\alpha=2s+1\;,\;\;\beta=s+1-2B_4\;,\;\;\gamma=s+1+2B_1\;,\nonumber
\\
&\delta=s+1+2B_2\;,\;\;\epsilon=s+1+2B_3\;,\;\;a=z_r\;,\nonumber
\\
&q=\dfrac{6s-3{\tilde{\lambda}}+(2s+1)(s+1)(r_+r_-+r_+^\prime r_-^\prime) \Lambda}{(r_- - r_-^\prime) (r_+- r_+^\prime) \Lambda} \nonumber\\
&\quad+\dfrac{3i(2s+1)\!\left[2\,\omega\, r_+ r_- - eQ(r_+ + r_-)\right]}
         {(r_- - r_-^\prime) (r_- - r_+)(r_+ - r_+^\prime) \Lambda}\,.\label{eq-q}
\end{align}  
One may easily verify that the Fuchs relation given in Eq.~\eqref{eq:Fuchs-relation} is satisfied automatically for the parameters presented in the above.

Now we are in a position to solve the (general) Heun equation~\eqref{eq:Heun}, in terms of the Heun functions. In general, by employing the Frobenius method, one may obtain two local solutions for a second order linear differential equation at each regular singular point. These local solutions include one regular and one with a branch cut; and for the Heun equation, they are normally referred to as Heun functions~\cite{Ronveaux1995,DLMF312}. We introduce $\mathrm{H\ell}(a, q; \alpha, \beta, \gamma, \delta; z)$ as the regular solution at $z=0$ with the normalization $\mathrm{H\ell}(a, q; \alpha, \beta, \gamma, \delta; 0)=1$.

In order to look for QNMs, one has to impose physically relevant boundary conditions. For this purpose, we shall first obtain the asymptotic solutions at each boundary. At $z=0$ ($i.e.$ the event horizon), the local solutions may be written as
%\begin{equation}
\begin{align}
y_{01}(z) = \,&\mathrm{H\ell}(a,\, q;\, \alpha,\, \beta,\, \gamma,\, \delta;\, z)\;,\nonumber\\
y_{02}(z) = \,&z^{\,1-\gamma}\,\mathrm{H\ell}\big(a, (a\delta+\epsilon)(1-\gamma)+q; \alpha+1-\gamma,\, \nonumber\\\qquad&\beta+1-\gamma,\, 2-\gamma,\, \delta;\, z\big)\;,\label{z=0}
\end{align}
%\end{equation}
where, for a non-integer $\gamma$, the Heun functions $y_{01}(z)$ and $y_{02}(z)$ are linearly independent. Similarly, the local solutions at $z=1$ ($i.e.$ the cosmological horizon) are
%\begin{equation}
\begin{align}
y_{11}(z) = \,&\mathrm{H\ell}(1-a,\, \alpha\beta-q;\, \alpha,\, \beta,\, \delta,\, \gamma;\, 1-z)\;,\nonumber\\
y_{12}(z) = \,&(1-z)^{\,1-\delta}\,\mathrm{H\ell}\big(1-a, ((1-a)\gamma+\epsilon)(1-\delta)\nonumber\\\qquad&+\alpha\beta-q;\alpha+1-\delta,\, \beta+1-\delta,\, 2-\delta,\, \gamma;\, 1-z\big)\;,\label{z=1}
\end{align}
%\end{equation}
where, for a non-integer $\delta$, the Heun functions $y_{11}(z)$ and $y_{12}(z)$ are linearly independent. For each case shown in the above, the convergence domain of the local solutions is a disk centered at the expansion point, with a radius equal to the distance to the nearest singularity.

In a linear system, the local solutions at different regular singular points may be connected linearly. For our case, therefore, the connection relations between the local solutions at $z=0$ and those at $z=1$ may be written as 
%\begin{equation}
\begin{align}
y_{01}(z) &= c_{11} y_{11}(z) + c_{12} y_{12}(z)\;,\nonumber\\
y_{02}(z) &= c_{21} y_{11}(z) + c_{22} y_{12}(z)\;,\label{eq:conn}
\end{align}
%\end{equation}
where we assume that $y_{11}(z)$ and $y_{12}(z)$ are independent. Note that here the coefficients $c_{ij}$ may be easily expressed in terms of the Wronskian
\begin{equation}
W[y_a,y_b] =
\begin{vmatrix}
y_a(z) & y_b(z) \\[0.9ex]
y_a'(z) & y_b'(z)
\end{vmatrix}\;,
\label{eq:wronskian_def}
\end{equation}
where $\prime$ denotes a derivative with respect to $z$. In asymptotically dS spacetimes, quasinormal frequencies are defined as \textit{ingoing} wave boundary conditions at the event horizon and \textit{outgoing} wave boundary conditions at the cosmological horizon. Based on the relation between the radial function $R_s$ and $y$ given in Eq.~\eqref{eqfactor} and by considering the Heun functions given in Eqs.~\eqref{z=0} and~\eqref{z=1}, one immediately realize that $y_{02}$ represents an ingoing wave solution at the event horizon whereas $y_{11}$ denotes an outgoing wave solution at the cosmological horizon. This indicates $c_{22}=0$, $i.e.$
\begin{equation}
c_{22}(\omega) = \frac{W[y_{02}, y_{11}]}{W[y_{12}, y_{11}]} = 0\;.
\label{eq:C22}
\end{equation}
Based on the definition of the Wronskian given in Eq.~\eqref{eq:wronskian_def} and by employing the built-in (general) Heun function \texttt{HeunG} in \textsc{Mathematica}, one may obtain quasinormal spectral of charged scalar and Dirac fields on RN-dS BHs with a global monopole.
%%%%%%%%%%%%%%%%%%%%%%%%%%%%%%%%%%%%%
\section{Numerical results}
\label{secresults}
%%%%%%%%%%%%%%%%%%%%%%%%%%%%%%%%%%%%%
In this section, by taking a few selected numerical results, we illustrate the impact of the global monopole on three different families of modes and on the implications of SCC, by using the \textit{Heun function method} outlined in Sec.~\ref{secnumerics}, for charged scalar and Dirac fields on RN-dS BHs with a global monopole. In numerics we take $M=1$ to measure all other physical quantities and introduce $n$ to represent the overtone number. 

\begin{figure*}[t]
	\centering
	\includegraphics[width=.9\textwidth]{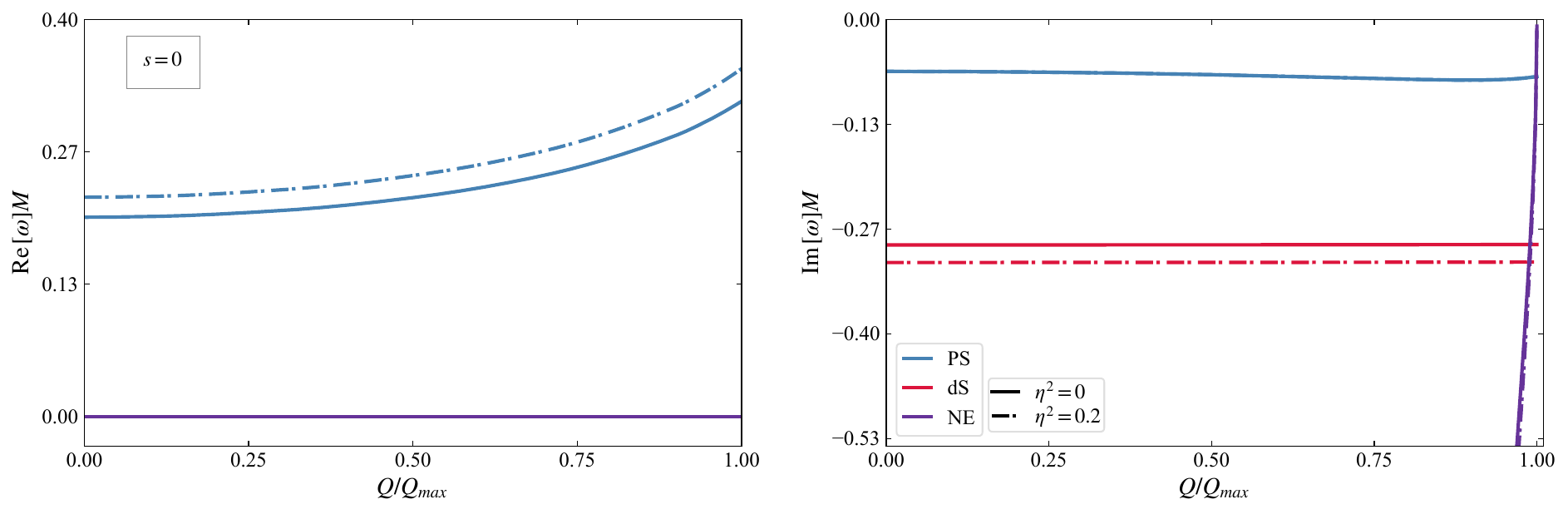}\\[0.9em]
	\includegraphics[width=.9\textwidth]{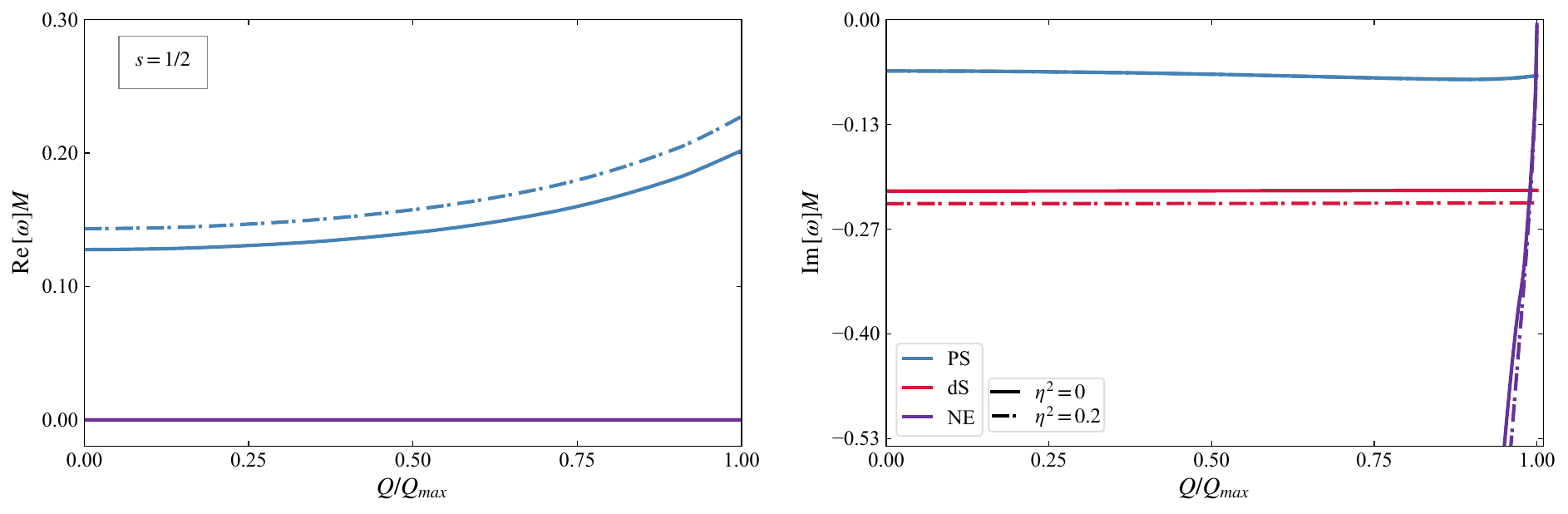}
	\caption{Real (left) and imaginary (right) parts of the fundamental quasinormal frequencies for scalar (top) and Dirac (bottom) fields in terms of $Q/Q_{\textit{max}}$, with $e=0$ and $\ell=1$ ($\ell=1/2$) for scalar (Dirac) fields. Note that here the blue, red and purple curves correspond to the PS, dS and NE modes, and the solid (dash-dotted) lines stand for $\eta^2=0$ ($\eta^2=0.2$).}
	\label{fig:fig1}
\end{figure*}

Before we proceed, we shall first verify the validity of the Heun function method. To this end, by taking $e=0$, $n=0$ and $\ell=1$ ($\ell=1/2$) for scalar (Dirac) fields as an example, we compute the corresponding quasinormal spectra in Fig.~\ref{fig:fig1}, by varying the background charge $Q$ both for $\eta^2=0$ and for $\eta^2=0.2$. For the case of $\eta^2=0$, we find an excellent agreement between our results generated by the Heun function method and the counterparts reported in the literature obtained by other numerical methods~\cite{Destounis:2018qnb,Cardoso:2017soq,Yoshida:2010zzb}; while for the case of $\eta^2=0.2$, we implement both the Heun function method and the matrix method adapted from our previous works~\cite{Lei:2021kqv}, and we obtain consistent results. These comparisons indicate that the Heun function method is a reliable approach to explore QNMs in asymptotically dS BHs. 

In addition, we make the following observations for the Heun function method. In numerics the matching point is chosen in the region $z \in (0,1)$ and one typically takes $z = 1/2$. Then we find, for near-extremal BHs characterized by $Q/Q_{\textit{max}} > 1 - 10^{-3}$, that the NE family becomes increasingly sensitive to the choice of initial values when the multipole number is moderately large ($e.g.$, $\ell > 10$), indicating that the efficiency of the method is reduced in this regime. This situation is somewhat improved for the dS family, although similar sensitivity may still arise. In contrast, within the parameter space explored here, the PS family does not exhibit this issue. The aforementioned restrictions, however, may be largely relaxed by choosing another matching point close to unity. As we have checked, by taking for example the matching point $z=99/100$, the sensitivity of quasinormal frequencies on the initial value may be significantly reduced, and the overall computation time is also shortened. By following such procedures, in numerics we are able to obtain modes with $\ell > 40$ for generic configurations and $\ell > 20$ for the near-extremal case with $Q/Q_{\max} > 1 - 10^{-3}$. These observations imply that the Heun function method is both accurate and efficient to compute the quasinormal spectra in asymptotically dS BHs.
%%%%%%%%%%
%%%%%%%%%%

Now we are ready to illustrate the monopole effects both in the context of QNMs and in the context of SCC.
%%%%%%%%%%%%%%%%%%
\subsection{Quasinormal spectra}
\label{secQNMs}
%%%%%%%%%%%%%%%%%%
In this subsection, we thoroughly explore the monopole effects on three different families of modes.

%%%%%%%%%%%%%%%%%%
%\subsubsection{Photon sphere modes}
\subsubsection{PS modes}
%%%%%%%%%%%%%%%%%%
In asymptotically dS BHs, a family of QNMs is associated with the dynamics of the photon sphere~\cite{Berti:2009kk}; these modes are therefore refereed to as the PS modes and are denoted by $\omega_{\mathrm{ps}}$. In this case, the effective potential of perturbation fields develops a single barrier outside the event horizon, whose maximum lies close to the unstable circular null geodesics. Waves temporarily trapped near this peak undergo damped oscillations before leaking either into the BH or out to the cosmological horizon. 

In the eikonal limit ($\ell \gg 1$), the correspondence between the PS modes and null geodesics becomes particularly transparent: the oscillation frequency is determined by the orbital angular velocity $\Omega_{\mathrm{ph}}$, while the decay rate is governed by the Lyapunov exponent $L_{\mathrm{ph}}$ of the unstable orbit~\cite{Cardoso:2008bp}. These modes are primarily dictated by the background geometry and are thus only weakly sensitive to the specific perturbing field. Following the rationale developed in~\cite{Casals:2020uxa}, in the eikonal limit and in the presence of a global monopole, the PS mode is given by
\begin{equation}
\omega_{\mathrm{ps}} = \Omega_{\mathrm{ph}} - i\!\left(n+\tfrac{1}{2}\right)L_{\mathrm{ph}}\;,
\label{wps}
\end{equation}
where
\begin{equation}
\begin{aligned}
\Omega_{\mathrm{ph}} &= \frac{\bigl({\tilde{\ell}}+\frac{1}{2}\bigr)\sqrt{\Delta_{\mathrm{ph}}} + e Q r_{\mathrm{ph}}}{r_{\mathrm{ph}}^{2}} \;,\\[3pt]L_{\mathrm{ph}} &= 2 \sqrt{\frac{ r_{\mathrm{ph}}\!(\Delta'_{\mathrm{ph}})^{2}+ 2\,\Delta_{\mathrm{ph}}\bigl(\Delta'_{\mathrm{ph}}-r_{\mathrm{ph}}\Delta''_{\mathrm{ph}}\bigr)}{ r_{\mathrm{ph}}\!\left(4r_{\mathrm{ph}}^{2}+r_{\mathrm{ph}}\Delta'_{\mathrm{ph}}-4\Delta_{\mathrm{ph}}\right)^{2}}}\;.
\end{aligned}
\label{eq:ph}
\end{equation}
Here \(r_{\mathrm{ph}}\) denotes the radius of the circular photon orbit, \(\Delta_{\mathrm{ph}} \equiv \Delta_r(r_{\mathrm{ph}})\), $\tilde{\ell}$ is the modified multipole number defined in Eq.~\eqref{modifiedangular}, and $\prime$ denotes a derivative with respect to $r$. Note that Eq.~\eqref{wps} provides good initial values to look for PS modes in our numerical calculations.

\begin{figure*}[t]
	\centering
	\includegraphics[width=.9\textwidth]{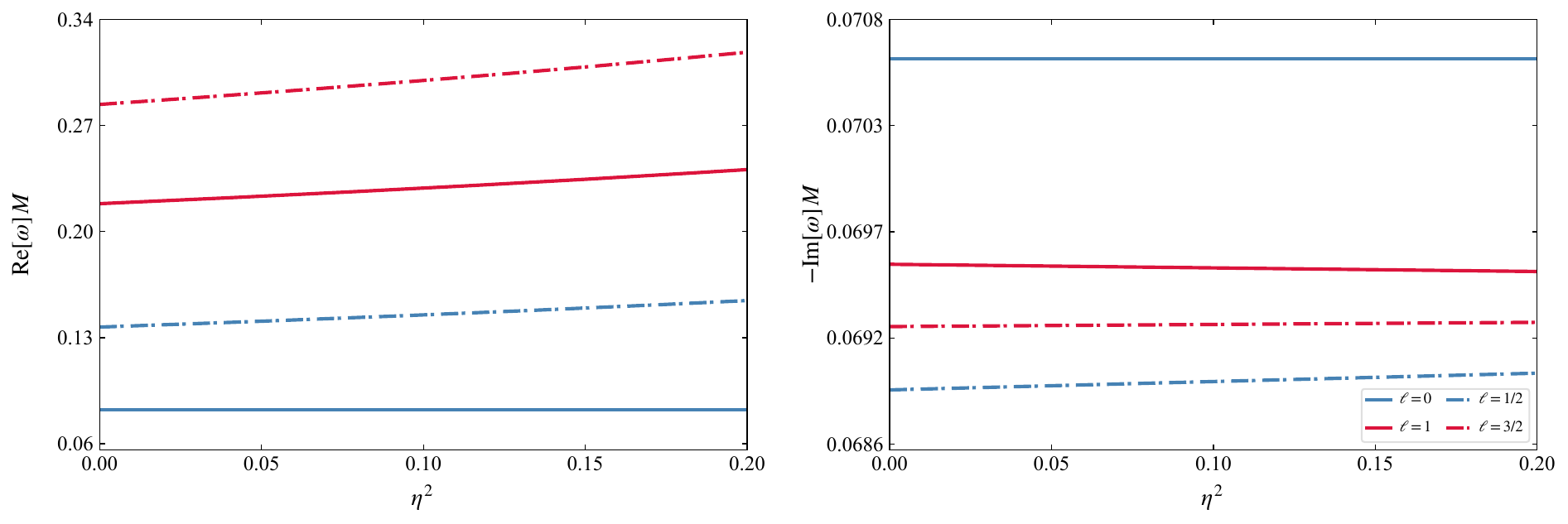}
	\caption{Real (left) and imaginary (right) parts of the fundamental \textit{PS} frequencies for scalar (solid) and Dirac (dash-dotted) fields in terms of the monopole parameter $\eta^2$, with $e=0$, $Q=0.5$ and $\Lambda=0.06$. Note that here we have taken $\ell=0$ (blue) and $\ell=1$ (red) for scalar while $\ell=1/2$ (blue) and $\ell=3/2$ (red) for Dirac fields.}
	\label{fig:PS-1}
\end{figure*}

\begin{figure*}[t]
	\centering
	\includegraphics[width=.9\textwidth]{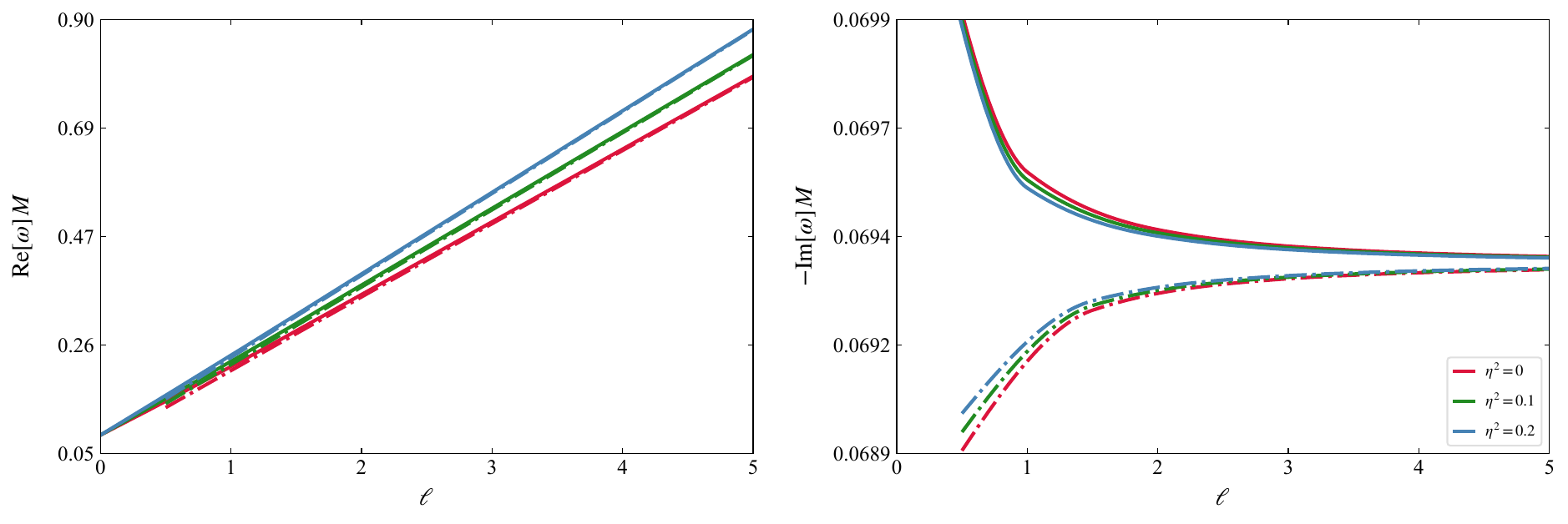}
	\caption{Real (left) and imaginary (right) parts of the fundamental \textit{PS} frequencies for scalar (solid) and Dirac (dash-dotted) fields in terms of the multipole number $\ell$, with the same parameters as in Fig.~\ref{fig:PS-1}. Note that here we have taken various values of the monopole parameter, including $\eta^{2} = 0$ (red), $0.1$ (green), and $0.2$ (blue).}
	\label{fig:PS-2}
\end{figure*}

We explore the effect of the global monopole on the PS modes, with fixed $Q = 0.5$, $\Lambda = 0.06$, $e=0$ and $n=0$, by varying the monopole parameter $\eta^2$ in Fig.~\ref{fig:PS-1} and by varying the multipole number $\ell$ in Fig.~\ref{fig:PS-2}. 

For all cases considered herein, except for $\ell=0$, we observe in Fig.~\ref{fig:PS-1} that the real part of the quasinormal spectra increases for both fields, while the magnitude of the imaginary part decreases for scalar but increases for Dirac fields, as the monopole parameter increases. In the exceptional case of $\ell=0$, the global monopole leaves the PS spectra unchanged. These trends may be understood by the influence of the multipole number on the PS modes. To demonstrate this fact, we present the impact of the multipole number on the PS frequencies in Fig.~\ref{fig:PS-2} and the effect of the monopole on the modified multipole number $\tilde{\ell}$ in Fig.~\ref{fig:ellvseta}. We find in Fig.~\ref{fig:PS-2} that, as the multipole number increases, the real part of the PS frequencies increases both for scalar and for Dirac fields, while the magnitude of the imaginary part increases for scalar but decreases for Dirac fields. In Fig.~\ref{fig:ellvseta} we display that $\tilde{\ell}$ increases monotonically in terms of $\eta^2$. Together, these results indicate that the global monopole alters the spectra through modifying $\ell$, which is exactly what we observed in Fig.~\ref{fig:PS-1}.

Moreover, in the eikonal regime, we find in Fig.~\ref{fig:PS-2} that, for both fields studied here, as $\ell$ increases, the real part of the PS modes scales linearly with $\ell$ while the imaginary part approaches a constant value. This also implies that the dominant mode occurs as $\ell \to \infty$ for scalar and at $\ell = 1/2$ for Dirac fields. Furthermore, as $\ell \to \infty$, the PS modes of scalar fields are the same with the counterparts of Dirac fields.
%%%%%%%%%%
%%%%%%%%%%
%%%%%%%%%%
%%%%%%%%%%
\subsubsection{dS modes}
%%%%%%%%%%
In a pure dS universe, perturbations of scalar and Dirac fields are governed by a set of purely imaginary characteristic frequencies. Following the procedures developed in~\cite{Du:2004jt,Lopez-Ortega:2006aal,Cardoso:2017soq,Casals:2020uxa}, when a global monopole is present, we have
\begin{equation}
\omega_{\text{dS}} = - i (\tilde{\ell} + n + 1) \kappa_{c}^{\text{dS}}\;,\label{dSmodes}
\end{equation}
where the surface gravity of the cosmological horizon is $\kappa_{c}^{\text{dS}} = \sqrt{\Lambda/3}$ and $\tilde{\ell}$ is, again, the modified multipole number defined in Eq.~\eqref{modifiedangular}. 

In the presence of a BH in a dS spacetime with a global monopole, the modes given in Eq.~\eqref{dSmodes} become a distinct family of QNMs, typically refereed to as the dS modes. In contrast to the PS modes, the decay rate of dS modes is primarily governed by the surface gravity of the cosmological horizon rather than by the photon sphere potential barrier. Moreover, Eq.~\eqref{dSmodes} is a good approximation for locating the dS modes in our numerical calculations.

\begin{figure*}[t]
	\centering
	\includegraphics[width=.9\textwidth]{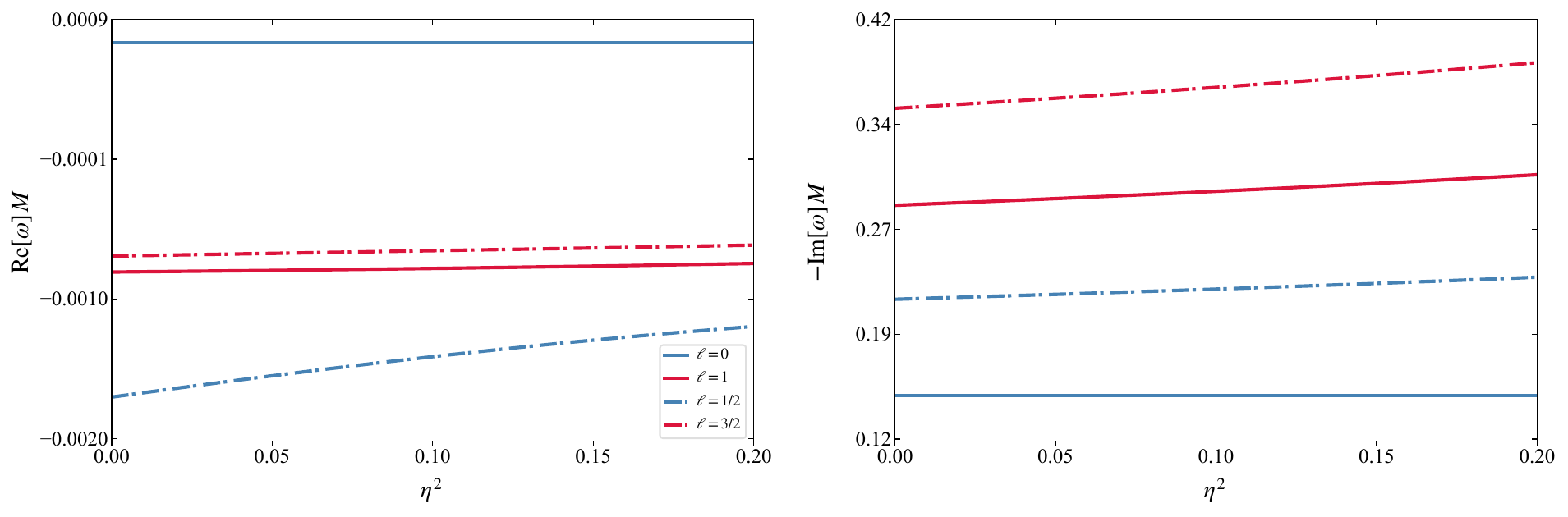}
	\caption{Real (left) and imaginary (right) parts of the fundamental \textit{dS} frequencies for scalar (solid) and Dirac (dash-dotted) fields in terms of the monopole parameter $\eta^2$, with $e=0.1$, $Q=0.5$ and $\Lambda=0.06$. Note that here we have taken $\ell=0$ (blue) and $\ell=1$ (red) for scalar while $\ell=1/2$ (blue) and $\ell=3/2$ (red) for Dirac fields.}
	\label{fig:dS-1}
\end{figure*}

\begin{figure*}[t]
	\centering
	\includegraphics[width=.9\textwidth]{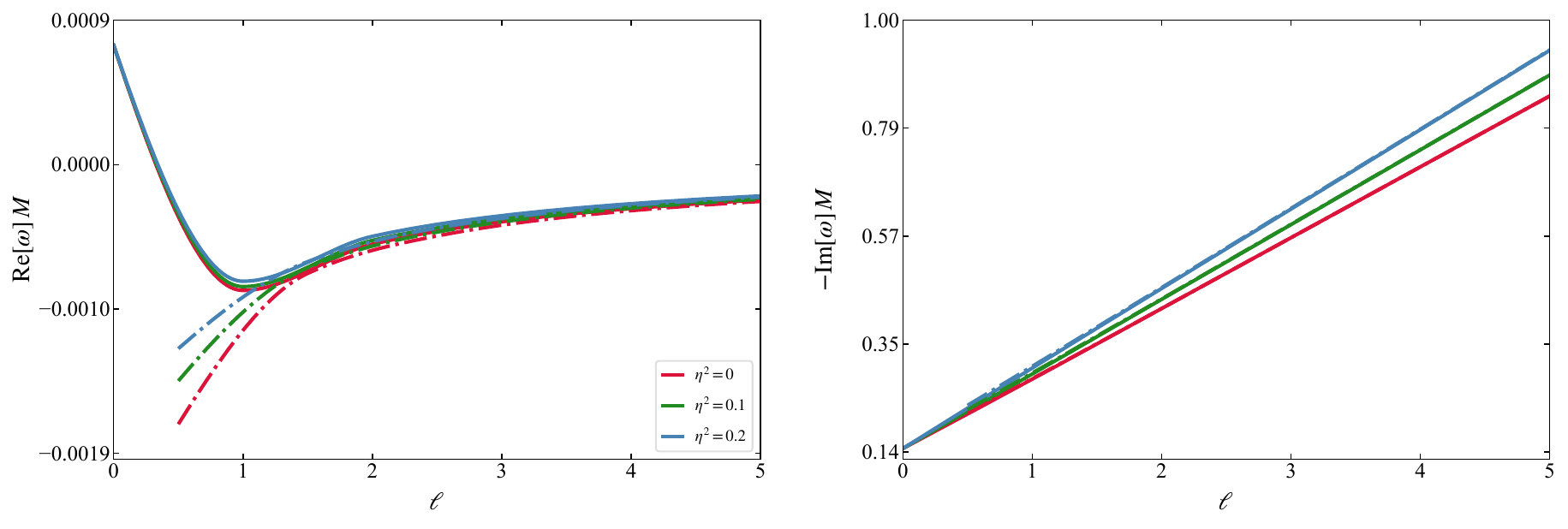}
	\caption{Real (left) and imaginary (right) parts of the fundamental \textit{dS} frequencies for scalar (solid) and Dirac (dash-dotted) fields in terms of the multipole number $\ell$, with the same parameters as in Fig.~\ref{fig:dS-1}. Note that here we have taken various values of the monopole parameter, including $\eta^{2} = 0$ (red), $0.1$ (green), and $0.2$ (blue).}
	\label{fig:dS-2}
\end{figure*}

Following the analysis performed for the PS modes, we now investigate the effect of the global monopole on the dS modes, with fixed $Q = 0.5$, $\Lambda = 0.06$, $e=0.1$ and $n=0$, by varying the monopole parameter $\eta^2$ in Fig.~\ref{fig:dS-1} and by varying the multipole number $\ell$ in Fig.~\ref{fig:dS-2}. 

From Fig.~\ref{fig:dS-1}, for both fields and for all cases with $\ell\neq0$, one observes that both the real and the magnitude of the imaginary parts of dS modes increase as the monopole parameter $\eta^2$ increases. In the exceptional case of $\ell=0$, the global monopole does not alter the dS spectra. These trends may be understood as follows. First, from Appendix~\ref{app}, we notice that the global monopole alters the equations of motion by changing the multipole number from $\ell$ to $\tilde{\ell}$ and, therefore, $\tilde{\ell}$ plays the same role as $\ell$. And Fig.~\ref{fig:ellvseta} shows clearly that the modified multipole number $\tilde{\ell}$ increases monotonically with respect to $\eta^2$, indicating that the impact of the global monopole on the spectra is the same as that of the multipole number. Then we present the impact of the multipole number on the dS frequencies in Fig.~\ref{fig:dS-2}. It shows that, for scalar fields with $\ell\geq1$ and for Dirac fields with $\ell\geq1/2$, as $\ell$ increases, both the real and the magnitude of the imaginary parts of dS frequencies increase, which is exactly the impact of the global monopole on the dS spectra we observed in Fig.~\ref{fig:dS-1}.

In Fig.~\ref{fig:dS-1}, it also shows that, for both fields and all cases considered here, the magnitude of the real part is much smaller than that of the imaginary part, implying that the dS modes are effectively purely damped. In Fig.~\ref{fig:dS-2}, for both fields and in the eikonal limit, we further observe that, ($i$)the magnitude of the imaginary part instead of the real part scales linearly with $\ell$; ($ii$) the dS frequencies coincide with each other, and the real part depends on $\eta^2$ weakly whereas the slope of the imaginary part is determined by $\eta^2$. Moreover, Fig.~\ref{fig:dS-2} shows that the slowest decaying mode is the fundamental mode with $\ell=0$ ($\ell=1/2$) for scalar (Dirac) fields.
%%%%%%%%%%%%%%%%%%%%%%%%
%%%%%%%%%%%%%%%%%%%%%%%%
%%%%%%%%%%
\subsubsection{NE modes}
%%%%%%%%%%
When the Cauchy horizon $r_-$ is close to the event horizon $r_+$, a new family of QNMs emerges, which is typically refereed to as the NE modes~\cite{Dias:2018ufh}. In a near-extremal RN-dS spacetime, the separation between the Cauchy horizon and the event horizon shrinks rapidly, so that the surface gravity at the event horizon $\kappa_+$ is close to the surface gravity at the Cauchy horizon $\kappa_-$. For charged scalar and Dirac fields on the RN-dS spacetime with a global monopole, the leading behavior of the NE modes are characterized uniformly by~\cite{Cardoso:2017soq,Destounis:2018qnb},   
\begin{equation}
    \omega_{\mathrm{NE}} = \frac{eQ}{r_-} - i\big(\tilde{\ell} + n + 1 \big)\kappa_- 
    = \frac{eQ}{r_+} - i\big(\tilde{\ell} + n + 1 \big)\kappa_+ \;,
    \label{wne}
\end{equation}
where the surface gravities are defined as $\kappa_{\pm}=\tfrac{1}{2}\lvert f^\prime(r_{\pm})\rvert$ with $f(r)=\Delta_{r}/r^{2}$, and $\tilde{\ell}$ is the modified multipole number defined in Eq.~\eqref{modifiedangular}. Note that in the near-extremal limit, $\kappa_{\pm}$ approach zero, indicating that the NE modes are long-lived. In addition, Eq.~\eqref{wne} provides good initial values to look for NE modes in numerical calculations. 

\begin{figure*}[t]
	\centering
	\includegraphics[width=.9\textwidth]{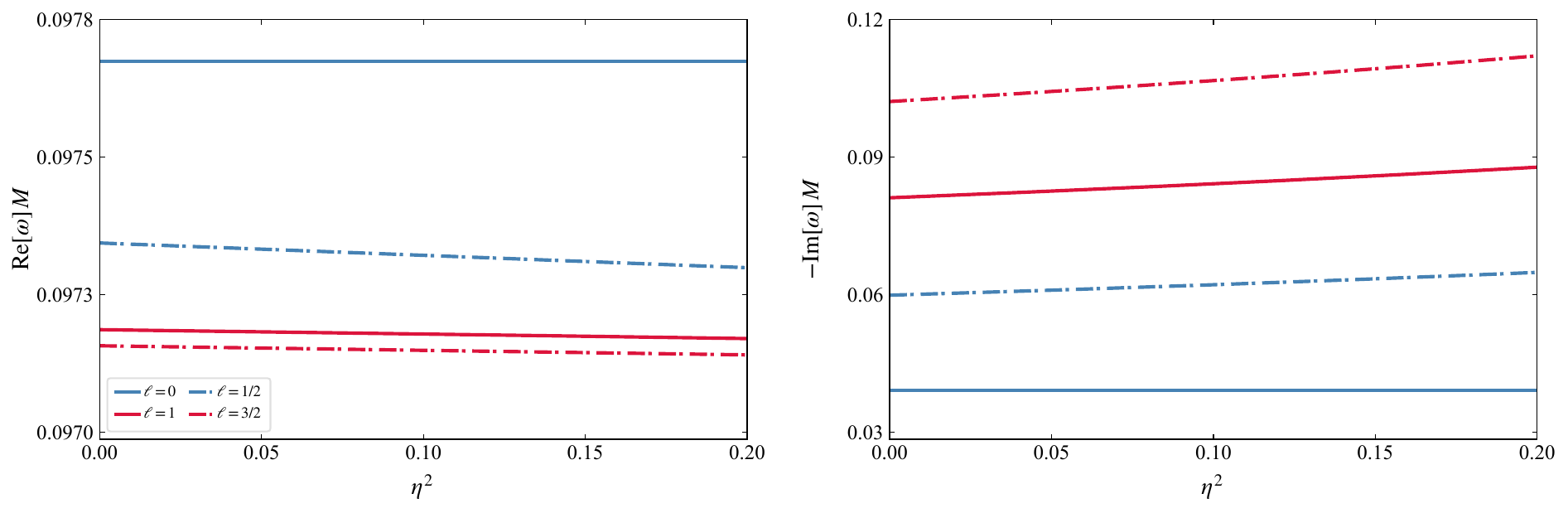}
	\caption{Real (left) and imaginary (right) parts of the fundamental \textit{NE} frequencies for scalar (solid) and Dirac (dash-dotted) fields in terms of the monopole parameter $\eta^2$, with $e=0.1$, $1-Q/Q_{\max}=10^{-3}$ and $\Lambda=0.06$. Note that here we have taken $\ell=0$ (blue) and $\ell=1$ (red) for scalar while $\ell=1/2$ (blue) and $\ell=3/2$ (red) for Dirac fields.}
	\label{fig:NE-1}
\end{figure*}

\begin{figure*}[t]
	\centering
	\includegraphics[width=.9\textwidth]{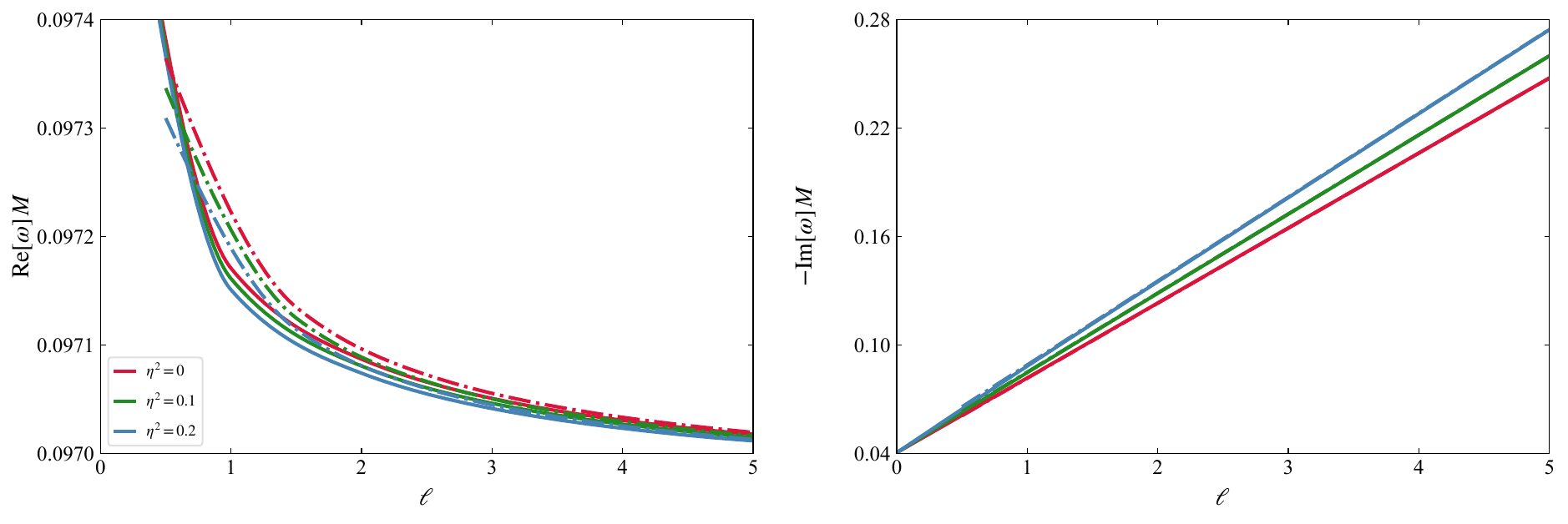}
	\caption{Real (left) and imaginary (right) parts of the fundamental \textit{NE} frequencies for scalar (solid) and Dirac (dash-dotted) fields in terms of the multipole number $\ell$, with the same parameters as in Fig.~\ref{fig:NE-1}. Note that here we have taken various values of the monopole parameter, including $\eta^{2} = 0$ (red), $0.1$ (green), and $0.2$ (blue).}
	\label{fig:NE-2}
\end{figure*}

Following the analysis of the PS and dS modes, here we start to investigate the effect of the global monopole on the NE modes, with fixed $1-Q/Q_{\max}=10^{-3}$, $\Lambda = 0.06$, $e=0.1$ and $n=0$, by varying the monopole parameter $\eta^2$ in Fig.~\ref{fig:NE-1} and by varying the multipole number $\ell$ in Fig.~\ref{fig:NE-2}. 

It shows in Fig.~\ref{fig:NE-1} that, for both fields and all cases with $\ell\neq0$, the real part of NE modes decreases whereas the magnitude of the imaginary part increases, as $\eta^2$ increases. For the exceptional case of $\ell=0$, the global monopole still leaves the NE spectra unchanged. Similarly to the previous two cases, since the monopole alters the spectra through the multipole number, one deduces that the impact of the global monopole on the NE spectra is the same as that of the multipole number. From Fig.~\ref{fig:NE-2}, we observe that, as $\ell$ increases, the real part of NE spectra decreases while the magnitude of the imaginary part increases; which gives exactly the trends induced by the monopole shown in Fig.~\ref{fig:NE-1}.

We also display in Fig.~\ref{fig:NE-2} that, for both fields, the real part of NE spectra depends on $\ell$ weakly whereas the magnitude of the imaginary scales (almost) linearly with $\ell$ and the corresponding slope increases as a function of $\eta^2$. In the eikonal regime, the NE modes of both fields become degenerate. We find, in particular, that the slowest decaying mode is the fundamental $\ell=0$ mode for scalar and the fundamental $\ell=1/2$ mode for Dirac fields. 
%%%%%%%%%%%%%%%%%%%%%%%%%%%%%%%%%%%%%%%%%%%%%%%%%%%%%%%
%%%%%%%%%%%%%%%%%%%%%%%%%%

%%%%%%%%%%%%%%%%%%%%%%%%%%%%%%%%%%%%%
\subsection{Strong Cosmic Censorship}
\label{secscc}
%%%%%%%%%%%%%%%%%%%%%%%%%%%%%%%%%%%%%
In this subsection, we move to investigate the impact of the global monopole on the implication of the SCC.

The SCC conjecture essentially states that the formed BH singularity is either spacelike or null, and the fate of SCC in BH spacetimes is closely related to the stability of the inner Cauchy horizon. In asymptotically flat charged or rotating BHs, late-time tails of perturbations have an inverse power-law decay behavior and undergo an infinite blueshift at the inner Cauchy horizon~\cite{Poisson:1989zz}. This triggers mass inflation and a strong curvature singularity, thereby supporting SCC~\cite{Ori:1991zz,Hamilton:2008zz}. 

This situation, however, changes dramatically when a positive cosmological constant is involved. In asymptotically dS spacetimes, the exterior decay of perturbations is exponential rather than polynomial and, as shown in~\cite{Cardoso:2017soq}, SCC may be violated in the near-extremal regime. In order to quantitatively measure the stability of the Cauchy horizon, and hence of the fate of SCC, one may introduce a dimensionless parameter $\beta$ with the definition~\cite{Hintz:2015jkj,Costa:2016afl}
\begin{equation}
\beta \equiv -\frac{\operatorname{Im}\!\left(\omega_0\right)}{\kappa_{-}}\;,\label{eq:beta}
\end{equation}
where $\omega_{0}$ stands for the nontrivial dominant quasinormal frequency and $\kappa_{-}$ is the surface gravity of the Cauchy horizon. If $\beta>1/2$, perturbations decay sufficiently slow so that the metric may be extended across the Cauchy horizon, thus violating SCC. Conversely, $\beta<1/2$ indicates that the Cauchy horizon remains unstable, thus preserving SCC. 

\begin{figure*}[t]
	\centering
	\includegraphics[width=.9\textwidth]{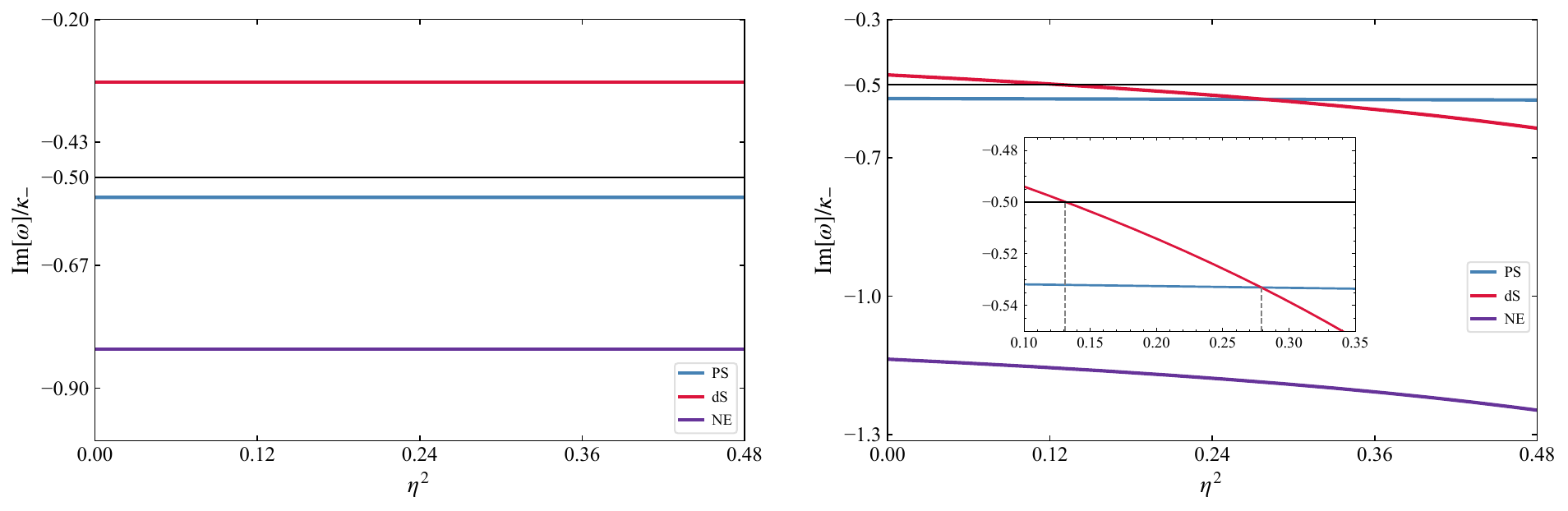}
	\caption{The impact of the global monopole on dominant modes of the PS (blue), dS (red) and NE (purple) types, with fixed $Q/Q_{\max}=0.992$, $\Lambda=0.008$, $e=0$ and $n=0$, both for scalar (left) and Dirac (right) fields. In the left panel for scalar fields, we have taken $\ell=5$ for PS, $\ell=0$ for dS and NE modes. In the right panel for Dirac fields, we have taken $\ell=1/2$ for all three types of modes.}
	\label{fig:SCC-1}
\end{figure*}

We investigate the impact of the global monopole on the implications of SCC in Fig.~\ref{fig:SCC-1}, by fixing $Q/Q_{max} = 0.992$, $\Lambda = 0.008$, $e=0$ and $n=0$, both for scalar (left panel) and for Dirac (right panel) fields. 

For scalar fields, we present all three families of modes with the lowest decaying rates and observe that they remain almost unchanged as $\eta^2$ is varied. This may be understood as follows. Based on the analysis in the previous subsection, we identify that the slowest decaying PS mode corresponds to the limit $\ell\rightarrow\infty$, whereas the slowest decaying dS and NE modes are the $\ell=0$ modes. Given that the monopole modifies the spectra by shifting the multipole number from $\ell$ to $\tilde{\ell}$, according to the relation between $\ell$ and $\tilde{\ell}$ in Eq.~\eqref{modifiedangular}, it becomes clear that, for the slowest decaying modes of all three types, the global monopole leaves them unaffected. Moreover, since the PS mode is dominant and satisfies the criterion $\beta<1/2$, the SCC is preserved. This indicates that the SCC remains respected even in the presence of a global monopole. We therefore conclude that, for scalar field perturbations, the fate of SCC is \textit{not} altered by the presence of a global monopole.

For Dirac fields, on the other hand, we again present all three families of modes with the lowest decaying rates. In contrast to the scalar case, we find that when $\eta^2\textless\;0.131$, the dS mode is dominant and the criterion $\beta<1/2$ is satisfied, implying that SCC is respected. As the monopole parameter increases and for $\eta^2\textless\;0.279$, the dS mode keeps dominant but the condition $\beta<1/2$ is violated, indicating that SCC is also violated. By further increasing the monopole parameter and when $\eta^2\textgreater0.279$, the PS mode overtakes the dS mode and becomes dominant, yet SCC remains violated. This demonstrates that, under Dirac perturbations, the violation of SCC is enhanced in the presence of a global monopole. 

\begin{figure*}[t]
	\centering
	\includegraphics[width=.9\textwidth]{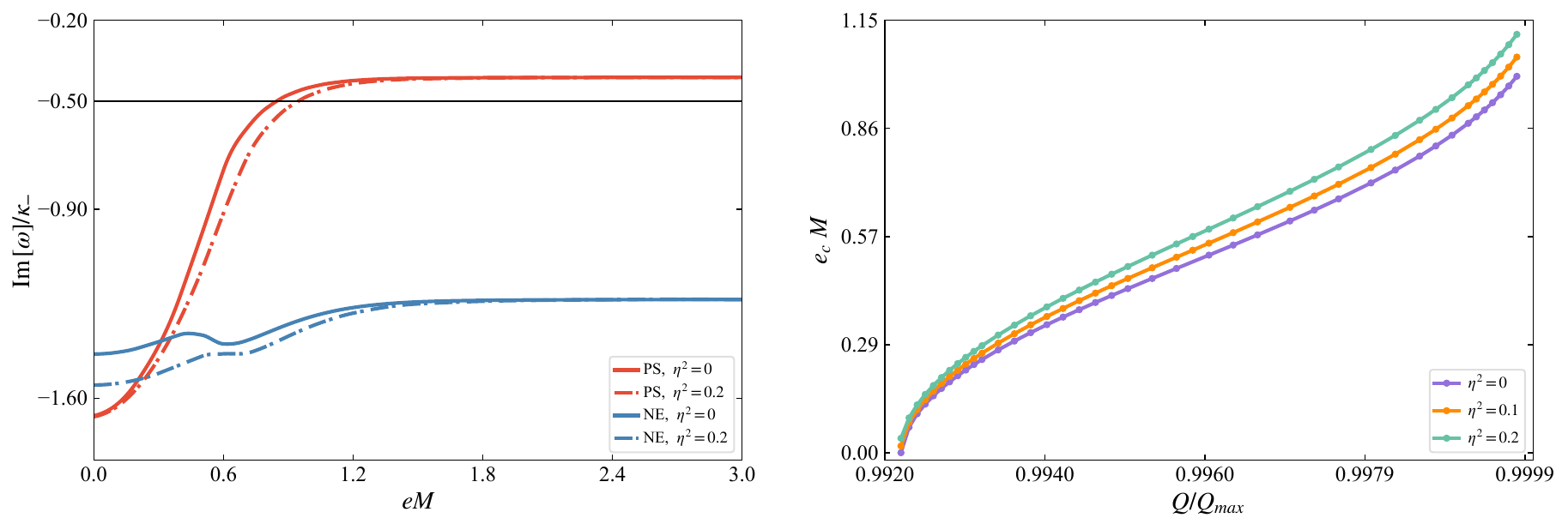}
	\caption{Left: The dominant modes of Dirac fields versus the field charge $e$, with fixed $1 - Q/Q_{\max} = 10^{-3}$, $\Lambda = 0.06$, $\ell=1/2$ and $n=0$, both for PS (red) and dS (blue) types. Here we take two representative global monopole parameters, $i.e.$ $\eta^{2}=0$ (solid) and $0.2$ (dash-dotted), and observe SCC holds for the case of $\eta^2=0$ ($\eta^2=0.2$) when $e \gtrsim 0.8445$ ($e \gtrsim 0.9443$). Note that the dS mode is absent in this regime. Right: The critical charge $e_c$ of Dirac fields versus the background charge $Q/Q_{\max}$ for the PS mode, with fixed $\Lambda = 0.06$, $\ell=1/2$, and $n=0$. Here we take three representative global monopole parameters, $i.e.$ $\eta^{2}=0$ (purple), $0.1$ (orange), and $0.2$ (green), as illustrations.}
	\label{fig:SCC-2}
\end{figure*}

By turning on the field charge, it has been argued that the violation of SCC may be mitigated~\cite{Mo:2018nnu}, and it has been shown for Dirac fields that there exists a critical field charge which marks the boundary between the parameter space respecting SCC and that violating it~\cite{Destounis:2018qnb}. Following the same logic, in Fig.~\ref{fig:SCC-2} we further explore, for various monopole parameters under Dirac perturbations, the charge effects on SCC in the left panel and the corresponding critical charge in the right panel, with fixed $1-Q/Q_{\max}=10^{-3}$, $\Lambda = 0.06$, $\ell=1/2$ and $n=0$.

In the left panel, it shows that, both for $\eta^2=0$ and for $\eta^2=0.2$, SCC is restored by increasing the field charge. Furthermore, it displays that a larger field charge is required to restore SCC when a global monopole is present compared to the case without it. This indicates that the violation of SCC is enhanced in the presence of a global monopole. In the right panel, the critical field charge $e_c$ for various monopole parameters is illustrated, showing that the critical charge increases as the monopole parameter increases. This finding further verifies the conclusion that the presence of a global monopole enhances the violation of SCC.  

Moreover, since the SCC may be recovered by charged field perturbations except in the case of highly charged BHs~\cite{Mo:2018nnu,Dias:2018ufh,Destounis:2018qnb}, and given that the Heun function method can be applied in such regimes as shown in Fig.~\ref{fig:SCC-2}, this further certifies that the Heun  function method is an efficient and robust approach for exploring the interaction between asymptotically dS BHs and perturbing fields. 
%%%%%%%%%%%%%%%%%%%%%%%%%%%%%%%%%%%%%
%%%%%%%%%%
%%%%%%%%%%
%%%%%%%%%%%%%%%%%%%%%%%%%%%%%%%%%%%%%%%%%%%%
%%%%%%%%%%%%%%%%%
\section{Discussion and Final Remarks}
\label{discussion}
%%%%%%%%%%%%%%%%%
In this paper, we have studied perturbations of charged scalar and Dirac fields on a global monopole RN-dS BH. To this end, we first derived the equations of motion both for scalar and for Dirac fields on the aforementioned background. For scalar fields with mass $\mu^2=\tfrac{2}{3}\Lambda$ and for massless Dirac fields, these equations were then reformulated uniformly into the Teukolsky equation. Since the Teukolsky equation in asymptotically dS spacetimes may be transformed into the (general) Heun equation, we have successfully obtained the quasinormal spectra and examined the SCC conjecture for both fields, by employing the Heun function method. In particular, we have addressed the impact of the global monopole on the spectra and on the implications of SCC.

In the context of QNMs, we have first explicitly shown that the Heun function method is an efficient and robust approach for computing quasinormal spectra in asymptotically dS spacetimes, not only for PS but also for dS and NE modes. Specifically, we have observed the following trends for the effects of the global monopole on these three types of modes when the multipole number $\ell\neq0$. For the PS modes, as the global monopole increases, the real part of the spectra increases for both fields while the magnitude of the imaginary part decreases for scalar but increases for Dirac fields. For the dS modes, on the other hand, both the real and the magnitude of the imaginary parts for both fields increase as the global monopole increases. In contrast to the above mentioned two cases, for the NE modes and for both fields, the real part decreases while the magnitude of the imaginary part increases as the global monopole increases. In the exceptional case of $\ell=0$ and for all three types, the global monopole leaves the spectra unaffected. 

In order to explore the impact of the global monopole on SCC, we first identified the dominant modes among three different types. For scalar fields, the dominant PS mode is governed by $\ell\rightarrow\infty$, whereas the dominant dS and NE modes are governed by $\ell=0$. In all these cases, the global monopole leaves the corresponding spectra unchanged. The SCC therefore remains unaffected by the presence of a global monopole under scalar field perturbations. This conclusion changes dramatically for Dirac fields. The dominant modes of Dirac fields for PS, dS and NE types are all governed by $\ell=1/2$. We found that the violation of SCC is enhanced due to the presence of a global monopole, for Dirac perturbations. 

With the definition of the modified multipole number $\tilde{\ell}$, the Teukolsky equation of a global monopole RN-dS BH was mapped into the standard Teukolsky equation of a RN-dS BH. Since $\tilde{\ell}$ plays the same role as $\ell$ and given the trends induced by $\ell$, we directly infer the role played by the global monopole in determining the spectra and the fate of SCC, as illustrated in the above.

By adding rotation to the background, equations of motion for scalar and Dirac fields are more involved. In order to investigate quasinormal spectra and the fate of SCC, one has to solve both the radial and the angular equations simultaneously. By performing such a study, it may help us not only to fully understand the impact of a global monopole both on the spectra and on SCC but also to further verify the robustness of the Heun function method in the sense that it may be employed to solve all three families of modes ($i.e.$ PS, dS and NE types) in a more general scenario. Work along this direction is underway and we hope to report on it in the near future.
\bigskip

%%%%%%%%%%%%%%%%%%%%%%%%%%%%%%%%%%%%%%%%%%%%%%%
\noindent{\bf{\em Acknowledgements.}}
%%%%%%%%%%%%%%%%%%%%%%%%%%%%%%%%%%%%%%%%%%%%%%%
This work is supported by the National Natural Science Foundation of China under Grant Nos. 12475050, 11705054, 12035005, by the Hunan Provincial Natural Science Foundation of China under Grant No. 2022JJ30367, by the Scientific Research Fund of Hunan Provincial Education Department Grant No. 22A0039, and by the innovative research group of Hunan Province under Grant No. 2024JJ1006.

\bigskip
\appendix
%%%%%%%%%%%
\section{The standard Teukolsky equation}
%\section{Equations of motion in the dimensionless form}
\label{app}
%%%%%%%%%%%
In this appendix, we illustrate that the Teukolsky equation derived on the \textit{global monopole} RN-dS BH, given in Eq.~\eqref{TeukolskyEq}, may be reformulated in the standard Teukolsky equation of a RN-dS BH, by refining a modified multipole number $\tilde{\ell}$.

To display this fact, by introducing the modified multipole number $\tilde{\ell}$ with the definition  
\begin{equation}
\tilde{\ell}=\dfrac{1}{2}\left(-1+\dfrac{\sqrt{(2\ell+1)^2+(\tilde{\eta}^2-1)(1-4s^2)}}{\tilde{\eta}}\right)\;,\label{modifiedangular}
\end{equation}
Eq.~\eqref{TeukolskyEq} turns into 
\begin{equation}
\begin{aligned}
\Bigg[&\, \Delta_r^{-s}\dfrac{d}{d r}\!\left( \Delta_r^{\,s+1}\frac{d}{d r} \right)+ \frac{K_r^{2}-isK_r\,\Delta'_{r}}{\Delta_r}+2isK_r^\prime\\\quad&-\dfrac{2\Lambda}{3}(s+1)(2s+1) r^{2}+2s - \bar{\lambda}\Bigg] R_s(r)=0\;,\label{TeukolskyEq2}
\end{aligned}
\end{equation}
where $\bar{\lambda}=\tilde{\ell}(\tilde{\ell}+1)-s(s-1)$, and Eq.~\eqref{TeukolskyEq2} is exactly the Teukolsky equation of RN-dS BHs, by replacing $\ell$ with $\tilde{\ell}$. This indicates that the impact of the global monopole on the spectra and on the implications of SCC is achieved by this modification, and our numerical results confirm that this is \textit{indeed} the case.
%%%%%%%%%%%
\begin{figure*}[t]
	\centering
	\includegraphics[width=.9\textwidth]{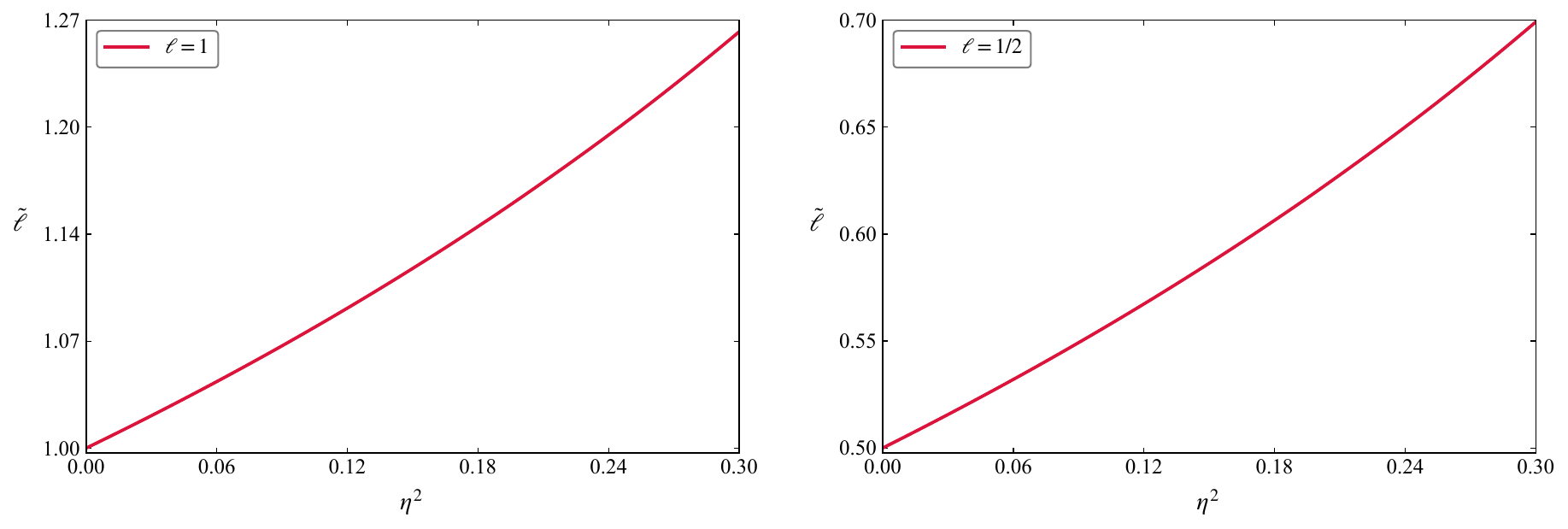}
	\caption{The modified multipole number $\tilde{\ell}$ in terms of the global monopole $\eta^2$, for scalar fields with $\ell=1$ (left) and for Dirac fields with $\ell=1/2$ (right). It shows clearly that $\tilde{\ell}$ increases monotonically as the monopole parameter $\eta^2$ increases.}
	\label{fig:ellvseta}
\end{figure*}
%%%%%%%%%%
\bibliographystyle{h-physrev4}
\bibliography{QNMsRNdS}

%\newpage

\end{document}